\documentclass[
aps,
final,
notitlepage,
oneside,
onecolumn,
nobibnotes,
nofootinbib,
superscriptaddress,
showpacs]
{revtex4}

\usepackage{graphicx}
\usepackage{amsmath}
\usepackage{amsfonts}
\usepackage{amssymb}
\usepackage{bbm}  
\usepackage{young}  
\usepackage{yfonts}  

\newcommand{\PLu}{\protect{Pl{\"u}cker}}
\newcommand{\PLL}{\protect{Pl{\"u}cker }}
\begin{document}
\title{On a Microscopic Representation of Space-Time VII -- On Spin}

%
\author{\firstname{Rolf}~\surname{Dahm}}
\email{dahm@bf-IS.de}
\affiliation{Permanent address: beratung f{\"{u}}r Informationssysteme und Systemintegration,
G{\"{a}}rtnergasse 1, D-55116 Mainz, Germany}
%

%

\begin{abstract}
We recall some basic aspects of line and line Complex representations,
of symplectic symmetry emerging in bilinear point transformations as 
well as of Lie transfer of lines to spheres. Here, we identify SU(2)
spin in terms of (classical) projective geometry and obtain spinorial
representations from lines, i.e.~we find a natural non-local geometrical
description associated to spin. We discuss the construction of a 
Lagrangean in terms of line/Complex invariants. We discuss the edges 
of the fundamental tetrahedron which allows to associate the most real
form SU(4) with its various related real forms covering SO($n$,$m$), 
$n+m=6$.
\end{abstract}

\maketitle


\section{Introduction}
At this stage (see \cite{dahm:MRST5} and references), we've 
established the ties from our original Lie group based approach 
of hadronic (and quantum) degrees of freedom in terms of SU(4)
and SU$*$(4), respectively, to (classical and advanced) projective
geometry of real 3-space. Now, in parts VII and VIII of this 
series, we want to focus more detailed on two of the included 
major aspects or building blocks -- spin and relativity. So 
departing from a description common to both parts and based 
on elementary line geometry, we are going to present a more 
detailed treatment of each spin and relativity related to 
classical line and projective geometry. This, however, at 
the same time almost immediately and inherently comprises 
a necessary branch of the two topics here and in upcoming 
papers. 
 
It is the task of this paper to recall briefly some (very)
elementary geometrical properties and features, and to 
relate this approach to spin and associated algebraical 
concepts and physical background. As such, we'll rearrange
coordinates and relate this rep\footnote{As in the papers 
before, we use the shorthand notation 'rep' to denote both
representations or realizations as long as the respective 
context is unambiguous. 'PG' subsequently denotes 'projective
geometry'.} to nowadays common spin (and 'quantum') aspects 
and some of their reps. Moreover, we use this approach 
to discuss further aspects as mentioned in the abstract
so that inherently we obtain a relation between the various
reps and, moreover, between their respective interpretations
and related concepts. Here, the crucial point with respect 
to spin discussions is the application of Lie's transfer 
principle \cite{lie:1872} and a special complexification 
of the real (inhomogeneous) coordinates.
On the other hand, this is also the branching point mentioned
above so that, based on the same Ansatz but on real coordinates,
in part VIII we'll concentrate on the original real (and the line
geometric) aspects, and there we'll continue this second branch
to discuss special linear Complexe and relativity.
We feel that the next few pages grant to state the relation 
of the usually disconnected 'quantum' world to classical
geometrical concepts, however, the wealth of possibilities
comprised in Lie's mechanism \cite{lie:1872}, e.g.~with 
respect to higher order (or class) objects or using further
choices of the transfer functions $f_{i}$, are still far 
from being applied or even exhausted. The 'degenerate cases'
mentioned in Lie's paper are related to very special choices
and cannot serve to comprise the full power of this Ansatz 
based on \PLu's algebraic concept of a general (higher 
dimensional) description of ordinary space and Lie's extension
thereof (see \cite{lie:1872}, especially page 151 above/end 
of \S 2 with respect to Lie's understanding of \PLu's original
algebraical Ansatz \cite{pluecker:1846}).

So although at this time we'll include general aspects only,
we feel it is the time to split our original series on a 
microscopic rep of space-time at this point into a spin (using 
complex numbers and elliptic geometry) and a relativistic 
(hyperbolic) part. On one hand, it is worth to investigate 
Lie transfer in more detail and to relate Clifford's and 
Study's work on elliptic geometry as well as especially 
Study's further work on (different) line geometries and on
kinematics (see e.g.~\cite{study:1903} and later publications).
To our opinion, in conjunction with (quaternary) invariant 
and form theory, this should be sufficient to exhaust and 
comprise what is nowadays called 'quantum theory'. On the
other hand, it is worth to separate a discussion focussing
on real aspects and hyperbolic geometry in order to keep 
in parallel the track to relativity - special relativity 
using (special) linear Complexe as well as the full geometry
of regular linear Complexe in $P^{5}$ and/or higher order
Complexe and nonlinear 
transformations. Stepping back to line and Complex geometry 
as the basic and underlying framework of Lie's approach 
\cite{lie:TransfGruppen}, we feel to escape the various 
restrictions of early use of (linear) rep theory of groups 
and algebras as well as concepts of differential geometry, 
and we think it's sufficient when asking for integral and/or
nonlinear representations only afterwards.

As such, in the upcoming sections, we use \PLu's line rep
(in inhomogeneous coordinates) and feel free to rearrange
and reinterpret this rep in a special form so that we can 
write the coordinates and transformations in matrix notation.
This helps to identify elements and relate them to quantum reps,
e.g.~in Lagrangean or current algebra notation, and to ongoing
discussions. Please note, too, that the next section is useful
both for spin and relativity, especially after switching to 
homogeneous coordinates and departing from affine to non-euclidean
geometry, and that it includes nothing but elementary PG. 
Afterwards, in section \ref{sec:lietransfer}, we branch from
real projective geometry by Lie's transfer principle \cite{lie:1872}
to the derivation of (complex) spin reps and the known algebraic
formalism. Throughout this section, we feel free to recall some
historic contexts and include them in the text, not to write 
some science-history related paper\footnote{According to current
discussions and the attribution of achievements, we feel the 
urgent necessity to reconsider the real historical development, 
however, personally, we neither have the competency nor the 
time to evaluate and honor those achievements scientifically 
and relate them in detail to the original authors. So we focus 
here on few spin-related aspects only which to our opinion need
(among other authors) to be attributed mainly to S. Lie and 
especially E. Study already back in the late $19^{th}$ and 
first decade of $20^{th}$ century instead of E. Cartan in 1938 
\cite{cartan:1938} (or even 1913, as he claimed there, see 
his introduction). However, we cannot rise a fundamental and 
historically complete priority discussion here.} but to put 
some known facts in an understandable context right from the
original development. The last section is devoted to a brief
outlook in order to sketch the wealth of opening possibilities
which appear by Lie transfer and sphere geometry.

\section{Line representation and coordinates}
The reason to cite some (very elementary) issues from 
\cite{plueckerNG:1868} is mainly to connect the reasoning
of the $19^{th}$ century view of projective geometry to 
the dynamic picture used nowadays in terms of 'time' and
'velocity' coordinates which emerge only after we tag 
points in the respective geometry and 'observe' them moving 
around\footnote{We want to emphasize the difference of the
'pictures' because there are alternative reps of points 
e.g.~by considering points in terms of (inter-)sections 
of lines in 3-space. This leads almost immediately to a
discussion of Congruences when including (moving) observers
into the physical description, see also our preparatory
discussions in \cite{dahm:MRST4}, \cite{dahm:MRST5}. We 
have to remark as well the {\it dependent} relation of 
velocities to the respective time coordinates of the 
individual lines, or axes, or of the (local) coordinate
systems which in its simplest form of (linear) motion 
with uniform velocity read as $s=v\cdot t$ in those 
coordinates. Note also, that \PLL (as well as Lie in 
his approach) uses (inhomogeneous) euclidean coordinates
$x, y, z$ and not fourfold/quaternary point coordinates, 
neither homogeneous nor (affine) Hesse coordinates by 
introducing the additional $t$-coordinate common at that 
time to denote the fourth (affine) point coordinate. This 
reduces the anharmonic ratio using point reps to a simple 
quotient of point distances or 'coordinate' projections.}. 
So in choosing \cite{plueckerNG:1868} the (euclidean) rep
\begin{equation}
\label{eq:basicline}
x\,=\,r\,z+\rho\,,\quad y\,=\,s\,z+\sigma\,,
\end{equation}
we obtain first of all a (cartesian) line rep of 3-space 
which is 'globally' defined by two linear functions in terms
of the third cartesian coordinate $z$. In each of the planes
$(x,z)$, i.e.~$y=0$, and $(y,z)$, i.e.~$x=0$, we find linear
line projections, too, which we may map to separate line reps
as well. So far, we do not have tagged special points on the 
line like the origin as a reference or a unit point or two 
(fundamental) points to declare anharmonic ratios or three 
points to declare a coordinate (or measure) on the line itself.
Moreover, in the same picture, we do {\it not yet} apply the
vector calculus of real 3-space common today by pointing with
a vector to the line and defining a 'time' parameter and a 
velocity on the line (or at least its direction).

\subsection{Coordinates}

So the following arguments are concerned with the (real)
parameters $r, \rho, s$ and $\sigma$ from eq.~(\ref{eq:basicline}), 
i.e.~what enters these equations are position and orientation 
(i.e.~'the geometry') of the line itself and as a 'whole' 
object in the 3-dim euclidean\footnote{We are going to extend 
this approach by homogeneous coordinates mainly in part VIII 
and postpone the more general discussion, also with respect
to anharmonic ratios and projective transformations of the
lines, to upcoming parts.} coordinate system $x,y,z$. Now 
\PLL founds his definition of (4-dim) line coordinates and 
their necessary extension\footnote{This ensures to preserve 
the grade during transformations whereas a sixth coordinate
is necessary to linearize the transformations of the line 
coordinates.} by a fifth line coordinate $\eta=r\sigma-s\rho$
on this basis, and in \cite{plueckerNG:1868} he shows how 
to define and use sets of lines in terms of Complexe, 
Congruences and Configurations as foundation of a space 
description, his so-called 'new geometry', in distinction
to standard (projective) geometry which uses the point 
as base element and planes to complete the point-based
framework with respect to duality (see e.g.~his contemporary
summary \cite{pluecker:1846} or later text books on PG 
also by other authors).

Suppressing PG details here, we want to prepare our discussion 
of both parts VII and VIII, i.e.~spin and relativity, by 
rearranging the (euclidean) coordinates $x,y$, still as a 
function of $z$ only, with respect to the line projection 
onto the $(x,y)$-plane at $z=0$. Our main concern and driving
force having been inclusion and understanding of special 
Lorentz transformations, however, it turned out that if we 
transform (in the point picture) along the $z$-axis, and if
we require as usual the $x$- and $y$-coordinates to remain
invariant, we have remarkable freedom in that this transform
concerns only $z\sim\tfrac{x_{3}}{x_{0}}$ as (euclidean) 
normal to $(x,y)$. Moreover, weakening the invariance 
requirement(s), we can rearrange the variables in the 
projection plane\footnote{We discuss more details related
to relativity in part VIII.} $(x,y)$ as needed to investigate
invariances of products, quotients, etc., too.

So if we arrange e.g.~the ratio of the two (cartesian) 
coordinate projections $x$ and $y$ of the line into a 
new coordinate $\zeta$ by $\zeta=\tfrac{x}{y}$, using 
eq.~(\ref{eq:basicline}) we obtain the relation 
$\tfrac{x}{y}=\zeta=\tfrac{r\,z+\rho}{s\,z+\sigma}$.
Now we can interpret the rhs as projective transformation\footnote{The
three (cartesian) projections of the original line onto 
each of the planes $(x,z)$, $(y,z)$ and $(x,y)$ in the 
general case are, of course, related to the line by
perspective.} and, of course, map the parameters to
a (real) matrix formalism by
\begin{equation}
\label{eq:euclideanProjTransf}
\frac{x}{y}=\zeta\longrightarrow
\left(
\begin{array}{c} x\\ y\end{array}
\right)\,=\,
\left(
\begin{array}{cc} r & \rho\\ s & \sigma\end{array}
\right)
\left(
\begin{array}{c} z\\ 1\end{array}
\right)\,.
\end{equation}

\subsection{Homogeneous Coordinates}
Moreover, by formally\footnote{Here, we may only use 
the affine (linear) coordinate extension by an absolute
plane $x_{0}=0$ (better by $\pm\sqrt{x_{0}^{2}}$) in 
the denominator.} introducing homogeneous coordinates 
for $\zeta\longrightarrow\tfrac{\zeta_{1}}{\zeta_{2}}$
and $z\longrightarrow\tfrac{x_{3}}{x_{0}}$, we can 
rewrite eq.~(\ref{eq:euclideanProjTransf}) according 
to
\begin{equation}
\label{eq:homogeneousProjTransf}
\left(
\begin{array}{c} \zeta_{1}\\ \zeta_{2}\end{array}
\right)
=
\left(
\begin{array}{cc} r & \rho\\ s & \sigma\end{array}
\right)
\left(
\begin{array}{c} x_{3}\\ x_{0}\end{array}
\right)
\end{equation}
and apply the usual $2\times 2$ matrix formalism.

Now, dependent on whether we focus on the rhs of 
eqns.~(\ref{eq:euclideanProjTransf}) and (\ref{eq:homogeneousProjTransf}),
or on the lhs of these equations, we steer towards a 
discussion of spin for the rhs or (special) relativity
for the lhs as we'll discuss soon. Note, however, that 
throughout our approach $\tfrac{x}{y}$ remains 'invariant'
if we replace the cartesian coordinates $x$ and $y$ 
homogeneously by $x\longrightarrow\tfrac{x_{1}}{x_{0}}$
and $y\longrightarrow\tfrac{x_{2}}{x_{0}}$, 
i.e.~$\tfrac{x}{y}=\tfrac{x_{1}}{x_{2}}$, following the
affine interpretation by tagging a certain plane within PG.

\PLu's fifth line coordinate $\eta=r\sigma-s\rho$ then turns
out to represent the determinant of the transformation matrix
above, i.e.~via the line rep in eq.~(\ref{eq:basicline}), we 
have as\-so\-ciated (formally) a projective transformation of
the (euclidean) $z$-axis to a (projective) transformation of 
the coordinate $\zeta$. Accordingly, we can relate this 'new'
coordinate axis of $\zeta$ to $z$ by associated projective
transformations $\zeta=f(z)$. More general, the projective
relation of $z$ and $\zeta$ allows to apply the full framework
of point transformations on first order (projective) elements
and to investigate binary forms (and the related invariant 
theory) in terms of mappings of point sets on lines to each 
other. In addition, our approach suggests the discussion of 
involutions on projective elements of first kind, so that 
we may include complex numbers via von Staudt's or L\"{u}roth's
geometric interpretation of complex numbers naturally. By 
considering this approach from a more general viewpoint, 
we have now set up the very foundation of the complete 
framework of PG, here in terms of point geometry on two 
lines coordinated by $z$ and $\zeta$.


\section{Lie Transfer and Spin}
\label{sec:lietransfer}
Following Lie's approach here, we are going to discuss 
the rhs of eq.~(\ref{eq:euclideanProjTransf}) only. Also, 
we require the matrix determinant (or \PLL coordinate 
$\eta$) here as nonzero, i.e.~the original line doesn't 
hit the $z$-axis, and the line of the respective projections
in the $(x,y)$-plane is skew versus the $z$-axis. In other
words, $(0,0)$ being the (cartesian) projection of the 
$z$-axis in the $(x,y)$-plane (and all its parallel planes,
i.e.~planes hitting the same 'absolute line' within the
spatial affine picture), the non-vanishing determinant 
ensures considering the case of two skew lines, and 
moreover eases the introduction of homogeneous coordinates
by avoiding singular/special cases. Some aspects related
to the general case, including $\eta=0$, will be discussed
in part VIII, also in the context of using homogeneous 
coordinates.

\subsection{Lie's Mapping -- aka 'Lie Transfer'}
Lie restricted the general cases of his mapping at first by 
considering lines instead of general curves (\cite{lie:1872},
\#5 and \#21), and further by choosing special constants 
in that the second Complex in the degenerate case (\cite{lie:1872},
\#21) hits the conic section in the absolute plane of the
{\it affine} geometry. Note, that it is this linear extension
of euclidean geometry\footnote{\ldots which also forces 
the switch to homogeneous coordinates to describe absolute 
elements. We refer to Study's remarks \cite{study:1907} 
and some of his further work cited here. See also our few 
remarks in section \ref{sec:studyswork}.} which introduces
the isotropy discussion (and as such the 'spinor' rep),
whereas with respect to relativistic (or non-euclidean) 
transformations, leaving a general second order 
(i.e.~quadratic) surface invariant, we have to switch to 
3-space instead of (linear) planar discussions. There will 
remain the connecting link that 2$^{nd}$ order surfaces 
can be generated by two families of lines, i.e.~in each 
point of the surface, we can find appropriate members of 
the two generating families, however, in order to handle
spin and (special) relativity from the viewpoint of the 
degeneracies discussed here and in the upcoming part VIII 
with respect to special linear Complexe, it is obvious 
that one should switch to Complexe in general as geometrical 
base elements as we have worked out and proposed throughout 
this paper sequence (see \cite{dahm:MRST5} and references 
there). A strongly supporting, direct argument for switching
to line and Complex geometry of $P^{5}$ would be to refer 
back to Lie's original investigations \cite{lie:1872} and 
by noting that the general theory underlying Lie's approach
and reasoning is Complex geometry. Indeed, Lie obtained 
this (bilinear) approach given in eq.~(\ref{eq:lietransfer})
only by further specializing his originally more general 
mapping to a line Complex hitting the absolute circle of 
affine geometry\footnote{It is obvious from the Cayley-Klein
construction of a metric that this construction is relevant
to metrical properties, too. However, we do not want to 
discuss general lengths, 'masses' or 'charges' here as this
approach is, first of all, connected to absolute elements 
of an affine Ansatz only.}, or isotropic lines, respectively.
For now, we think that Lie's Ansatz in general establishes 
sphere geometry in a separate space besides the line (and 
the represented Complex) geometry of ordinary 3-space, 
whereas concerning 'spin', here we treat only simple/linear
and (multiply) degenerate cases of Lie's general mapping 
which 'overlap' or relate in certain aspects due to the 
(linear) cases and/or the degeneracies considered.

\subsection{Matrix Reps and Spin}
\label{sec:matrixreps}
Nevertheless, we depart from Lie's restricted (bilinear) Ansatz
(see \cite{lie:1872}, p.~146, or \cite{lie:1872}, \S 8, \#24 (1)),
\begin{equation}
\label{eq:lietransfer}
-Zz=x -(X+iY)\,,\quad (X-iY)z=y-Z
\end{equation}
to relate the (cartesian) coordinates of the spaces $r$ and $R$,
which we rearrange into
\begin{equation}
\label{eq:lietransferrearranged}
x=-Zz+(X+iY) \,,\quad y=(X-iY)z+Z\,,
\end{equation}
and according to our choice\footnote{For detailed analytic 
consideration and calculations one should recall that at the
time of \PLu's and Lie's writing, the authors were used to
coordinate systems with different handedness, especially to
a different sign of the $y$-coordinate, which sometimes leads
to signs especially in linear equations involving $y$ or 
$\sigma$.} of the line representation in eq.~(\ref{eq:basicline})
and the matrix definition in (\ref{eq:euclideanProjTransf}),
this allows to identify the transformation matrix directly 
by
\begin{equation}
\label{eq:matrixrep}
\left(
\begin{array}{cc} r & \rho\\ s & \sigma\end{array}
\right)
\,\sim\,
\left(
\begin{array}{cc} -Z & X+iY\\ X-iY & +Z\end{array}
\right)
\,.
\end{equation}
Note here, that on the lhs, we originally describe a {\it linear}
transformation of {\it cartesian} coordinate projections which
yields the context for discussion of \cite{clebsch:1872} later 
on. So the original line coordinates of space $r$ are represented
(or 'replaced') by (complexified point) coordinate combinations 
of the space $R$, where the rhs of eq.~(\ref{eq:matrixrep}), 
however, may be re-expressed in nowadays common notation according
to
\begin{equation}
\label{eq:spin}
\left(
\begin{array}{cc} -Z & X+iY\\ X-iY & +Z\end{array}
\right)\,=\,
X\sigma_{1}-Y\sigma_{2}-Z\sigma_{3}
\end{equation}
when introducing standard 'Pauli matrices' $\sigma_{i}$. So 
the important point here -- derived from the background of 
Complex geometry -- is the identification of real (point) 
coordinates of $R$ in the Pauli/spin basis by eq.~(\ref{eq:spin})
with (inhomogeneous) line coordinates\footnote{This is especially
noteworthy facing the current discussions on non-local aspects
of quantum theories if we depart from real parametrizations 
of Pauli matrices. Note in this contexts, that already special
linear Complexe are {\it sets} of lines in 3-space, and that
using Complexe, we also have to take into account regular and
higher order Complexe, i.e.~the geometry of P$^5$ as well as
quadratic manifolds. Here, we are discussing linear mappings
and degenerate cases only, so these are only the very beginnings
of a potentially relevant description of observations in nature.}
of real 3-space, and it is the su(2) (Lie) algebra which maps 
the geometries and the respective coordinate interpretation. 
However, for the general environment and the context of our 
current discussion, it is necessary to mention and cite three
more important aspects given by Lie in his paper \cite{lie:1872}.

\subsection{Geometry in $R$-space}
\label{sec:geom-r-space}
Casting the equations above in a form to identify coordinates in
$R$-space, Lie shows (\cite{lie:1872}, p.~168) that those (still
inhomogeneous) coordinates fulfil the equation $R^{2}+S^{2}+1=0$,
or when cast in differential form\footnote{We cite the expression
as given, not discussing partial derivatives. Please note, that 
this replacement of coordinates by derivatives is attached to \PLu's
original line rep and to a {\it linear} choice of the coordinates.
So this Ansatz yields the very foundation of Lie's contact transformations 
based on line elements (or five-parameter space reps $(x,y,z,p,q)$)
later on. However, due to the intrinsically linear rep, we may also
switch directly to the point rep on lines in terms of binary forms
\cite{clebsch:1872} and the related symbolism.}, based on 
\cite{lie:1872}, \S 8, eq.~(3), by\footnote{We want to mention the 
relation to three homogeneous (planar) coordinates only.}
$R=\tfrac{\mathrm{d}X}{\mathrm{d}Z}$
,
$S=\tfrac{\mathrm{d}Y}{\mathrm{d}Z}$
, which yields $\mathrm{d}X^{2}+\mathrm{d}Y^{2}+\mathrm{d}Z^{2}=0$
and suggests the considerations of isotropic elements.
This led Lie to the statement that the line Complex in $R$-space
is being built of imaginary lines of zero length\footnote{Isotropic
lines.}. We want to address those aspects later and separately.

Here, to connect to 'Cartan's' spin rep, we want to recall Lie's
equations (\cite{lie:1872}, p.~168) of the line coordinates $R$ 
and $S$,
\begin{equation}
\label{eq:lie2cartan}
R=\frac{1}{2}\left(z-\frac{1}{z}\right)\,,\quad
S=\frac{1}{2i}\left(z+\frac{1}{z}\right)\,,
\end{equation}
and the quadratic equation $R^{2}+S^{2}+1=0$ from above. We can 
now start from either 'side' -- using the quadratic equation and
introducing homogeneous coordinates, or recalling the inhomogeneous
character of $z$ and introducing homogeneous quaternary coordinates
of 3-space in (\ref{eq:lie2cartan}). By the second approach, we 
obtain from eq.~(\ref{eq:lie2cartan})
\begin{equation}
\label{eq:lie2cartan2}
R=\left(\frac{z^{2}-1}{2z}\right)=\left(\frac{x_{3}^{2}-x_{0}^{2}}{2x_{0}x_{3}}\right)\,,\quad
S=\left(\frac{z^{2}+1}{2iz}\right)=\left(\frac{x_{3}^{2}+x_{0}^{2}}{2ix_{0}x_{3}}\right)\,.
\end{equation}
By introducing primed 'homogeneous' variables $R=:\tfrac{R'}{T'}$, 
$S=:\tfrac{S'}{T'}$, we can extract the identifications
\begin{equation}
\label{eq:lie2cartanhomog}
R'=x_{0}^{2}-x_{3}^{2}\,,\,
S'=i\left(x_{0}^{2}+x_{3}^{2}\right)\,,\,
T'=-2x_{0}x_{3}\,,
\end{equation}
where we've used the standard relation $z=\tfrac{x_{3}}{x_{0}}$
from affine geometry with respect to the $z$-axis in all places.
The equation $R^{2}+S^{2}+1=0$ converts to $R'^{2}+S'^{2}+T'^{2}=0$,
thus featuring (finite) line coordinates and 4$^{th}$ order 
surfaces in the original (point) coordinates. Eq.~(\ref{eq:lie2cartanhomog})
yields Cartan's coordinate and 'spinor' 'definition' in 
\cite{cartan:1938}, \#52.

\subsection{Lie Transfer and 'Spinors'}
Now regarding eqns.~(\ref{eq:lie2cartan2}) and (\ref{eq:lie2cartanhomog}),
it is easy to trace back Cartan's miraculous 'definition' of 
'a spinor' emerging in \cite{cartan:1938}, \#52 by recalling
the relation of inhomogeneous and homogeneous coordinates of
(real) 3-space, the definition of the absolute element of 
affine geometry. What enters in addition, are the absolute
circle, by $x_{1}^{2}+x_{2}^{2}+x_{3}^{2}=0$, $x_{0}=0$ in 
{\it homogeneous} spatial coordinates, and an identification 
$x_{0}^{2}-x_{3}^{2}$ here with $\xi_{0}^{2}-\xi_{1}^{2}$ 
in Cartan's notation. Accordingly, his 'spinor' coordinates
$\xi_{0}$ and $\xi_{1}$ map to two homogeneous point 
coordinates $x_{0}$ and $x_{3}$ of real 3-space, whereas by 
eqns.~(\ref{eq:lie2cartan2}) or (\ref{eq:lie2cartanhomog}) 
the three linear coordinates $x_{i}$ of Cartan map to Lie's
'line coordinates' $R$ and $S$, or $R', S'$ and $T'$ from 
above and from the 'inverse' Lie mapping, obeying 
$R^{2}+S^{2}+1=0$ or $R'^{2}+S'^{2}+T'^{2}=0$, respectively.

There is, however, a much easier way to establish the 'spinor'
interpretation versus Lie's mapping by starting from 
\cite{cartan:1938}, \#55, eq.~(1), and by recalling 
$z=\tfrac{x_{3}}{x_{0}}\sim\tfrac{\xi_{1}}{\xi_{0}}$ 
from above, i.e.
\begin{equation}
\xi_{0}x_{3}+\xi_{1}(x_{1}-ix_{2})=0\,,\,
\xi_{0}(x_{1}+ix_{2})-\xi_{1}x_{3}=0\,,
\end{equation}
Here, we can at first change the order of the equations
and divide by $x_{0}$, thus switching back to inhomogeneous,
cartesian coordinates $X$, $Y$ and $Z$. We obtain
\begin{equation}
-\xi_{1}Z+\xi_{0}(X+iY)=0\,,\,
\xi_{0}Z+\xi_{1}(X-iY)=0\,,
\quad\mathrm{or}
\end{equation}
\begin{equation}
\label{eq:help2}
0=-Z\xi_{1}+(X+iY)\xi_{0}\,,\,
0=(X-iY)\xi_{1}+Z\xi_{0}\,.
\end{equation}
We may put these equations in matrix form, too, by 
extracting a 'spinor' $(\xi_{1}, \xi_{0})^{T}$ to 
the right, and celebrating the su(2) (Lie) algebra 
in terms of Pauli matrices. What is more important,
however, is the comparison with eq.~(\ref{eq:lietransferrearranged})
if we introduce an inhomogeneous coordinate 
$\xi:=\tfrac{\xi_{1}}{\xi_{0}}$ (which, as we see 
above, is Lie's original $z$). Dividing eq.~(\ref{eq:help2}) 
by $\xi_{0}$, we obtain
\begin{equation}
\label{eq:cartanrewritten}
0=-Zz+(X+iY)\,,\,
0=(X-iY)z+Z\,,
\end{equation}
which recovers a special case of Lie transfer, i.e.~instead 
of Lie's general coordinate projections $x$ and $y$, Cartan's 
definition is obviously fixed to the $z$-axis at $(0,0)$
in the $(x,y)$-plane, which from Lie's approach is related
to vanishing $\eta$ (or determinant).

More important is the missing dependence of $x$ and $y$ 
at all, or according to (\ref{eq:cartanrewritten}) $x=0$ 
and $y=0$. So in addition to the restrictions and the 
resulting degeneracies/singularities applied by Lie, there
is {\it a priori no suitable coordinate dependence} besides
the 'spinor' itself describing a line in terms of two 
homogeneous coordinates $\xi_{i}$, or a simple 1-dim projective 
transformation of the quotient. The quotient may be reversed
by the exchange $\xi_{1}\longleftrightarrow\xi_{0}$, in 
'spinor' notation induced by $\sigma_{1}$ or $i\sigma_{2}$,
which correspond to reciprocal transformations\footnote{By 
a suitable setup of the system, arranging a (metric) radius
along $\hat{z}$, we have $r\longleftrightarrow\pm\tfrac{1}{r}$,
and planar duality as well. Conjugation of eq.~(\ref{eq:cartanrewritten})
in $R$-space is invariant if, at the same time, we perform
$z\longleftrightarrow-\tfrac{1}{z}$, etc. Note also sign 
behaviour of $R$ and $S$ in eq.~(\ref{eq:lie2cartan}) in
the $\pm$-cases. Such methods can be applied throughout 
Cartan's 'spinor' formalism, however, it should at least
be founded on Lie's approach. The more general treatment
has to include and respect polar systems.} $z\longleftrightarrow\pm\tfrac{1}{z}$.
Due to the remarkable situation that (in the usual point 
picture) special relativity doesn't transform $x$ and $y$
when we choose the transformation along $z$, Cartan's 
approach -- being independent of $x$ and $y$ -- survives
and reflects those transformations. One should, however, 
place question marks to general space-dependent SU(2)
spinors like $\psi(x)$ or $\psi(x^{\mu})$ founding on
Cartan's definition.

\subsection{Further Linear Aspects}
The second aspect covers a more general case. Here, Lie's
equations (\ref{eq:lietransfer}) mapping the two spaces 
$r$ and $R$, relate both spaces in a manner that points 
in $r$ map to isotropic lines in $R$, and points in $R$
map to the linear Complex $r+s=0$ in $r$. In upcoming 
papers, we hope to have the possibility to discuss more 
geometrical details, especially with respect to further
details and geometrical relations mentioned by Lie in 
\cite{lie:1872}. For now, we want to close this subsection
by reference to Lie's remarks in \S 9, p.~171 with respect
to the second unnumbered equation (see also eq.~(\ref{eq:Rsphere})) 
holding a quadratic relation. Tracing this back with 
our matrix rep, it is obvious that the variable $H$ in 
$R$ corresponds to the {\it unit matrix} $\sigma_{0}$, 
and it thus relates the unit matrix of the $R$ matrix 
rep with the linear Complex $r+\sigma$ of $r$-space and
it's special r\^{o}le. In Lie's original work \cite{lie:1872},
p.~171, the spheres in $R$-space (i.e.~using \PLu's inhomogeneous
coordinates\footnote{Generalizing the coordinates via 
affine homogeneous coordinate extensions, note that 
$\rho$ and $\sigma$ are related to the 'new' affine 
coordinate $t$ or $x_{0}$ and as such to the absolute 
plane in $r$-space. Recall also, that transformations 
of the inhomogeneous \PLL coordinates $r,s,\rho,\sigma$
do not preserve grade, i.e.~(\ref{eq:inhomogvars}) 
pretends a linear relationship due to the projective
rep.} to describe Complex lines) fulfil
\begin{equation}
\label{eq:Rsphere}
[X-(s+\rho)]^{2}+[Y-i(s-\rho)]^{2}+[Z-(\sigma-r)]^{2}=(\sigma+r)^{2}\,.
\end{equation}
So the coordinate set\footnote{Note, that $Y'$ is imaginary, 
or $iY'$ is real, and remember polar relationships in affine 
geometry, too.} $X'=s+\rho$, $Y'=i(s-\rho)$ and $Z'=\sigma-r$ describes 
the center of the sphere in $R$-space in terms of line coordinates 
(or a shift from an original center $(0,0,0)$) whereas $\sigma+r$ 
takes the r\^{o}le of an oriented coordinate or even a 'radius' 
$H=r+\sigma$ in terms of line coordinates of the line Complex in 
$r$-space. This enhances the mapping $r \longleftrightarrow -Z$, 
$\rho\longleftrightarrow X+iY$, $s \longleftrightarrow X-iY$ and 
$\sigma\longleftrightarrow +Z$ in eq.~(\ref{eq:matrixrep}) into (see 
\cite{lie:1872}, \S 9, eq.~(3))
\begin{equation}
\label{eq:inhomogvars}
r=\frac{1}{2}(\pm H'-Z')\,,
\rho=\frac{1}{2}(X'+iY')\,,
s=\frac{1}{2}(X'-iY')\,,
\sigma=\frac{1}{2}(\pm H'+Z')
\end{equation}
which now corresponds to the expression
\begin{equation}
\label{eq:maptopauli}
\frac{1}{2}\left(
\begin{array}{cc} -Z'\pm H' & X'+iY'\\ X'-iY' & +Z'\pm H'\end{array}
\right)
\sim
X'\sigma_{1}-Y'\sigma_{2}-Z'\sigma_{3}\pm H'\sigma_{0}
\,.
\end{equation}
Recalling \PLu's fifth coordinate $\eta=r\sigma-s\rho$, 
eq.~(\ref{eq:inhomogvars}) yields $r\sigma=H'^{2}-Z'^{2}$ and 
$s\rho=X'^{2}+Y'^{2}$, so $\eta=H'^{2}-Z'^{2}-X'^{2}-Y'^{2}$.
This prepares the basis of the 2-spinor formalism and all its
applications and analytical forms, in addition, we thus map
the quest for 'relativistic' invariance to 2$\times$2-matrix 
notation (\ref{eq:maptopauli}) or complexified quaternions, 
and invariance of the determinant $\eta$.

Especially, this gives some insight on how (due to the quadratic
nature of the manifold) the unit matrix serves \cite{lie:1872} in
this rep: A line given in $r$-space determines a sphere in $R$. 
However, a sphere in $R$ given by $(X,Y,Z,H^{2})$ maps to two 
lines in $r$, both being reciprocal polars with respect to the
linear line Complex $\pm H=r+\sigma=0$ in $r$. $H$ in $R$-space
plays the r\^{o}le of a sphere radius, so choosing $H=0$ amounts
to 'point spheres'\footnote{German: Punktkugeln \cite{lie:1872},
p.~171} mapping uniquely to the Complex $r+\sigma=0$ of $r$-space.

Finally, and pointing towards a third important aspect, it is 
worth mentioning \cite{lie:1872} article \#28 which relates the
contact of Lie's spheres to intersection of lines. This emphasizes
the relevance of contact interactions in both pictures when 
constructing and investigating symmetry and invariance principles.

For us, the essence of the reasoning so far -- remembering all 
the various restrictions applied up to this point -- justifies
according to our current use of the Pauli spin (or the quaternion)
algebra in various physical contexts to reverse the viewpoint: 
If we are to interpret Pauli matrices with {\it real} coefficients
and including the unit matrix to represent a 'radius', we have 
to recall their origin from transformations in $R$-space and 
their identification with some (inhomogeneous) line coordinates
which themselves describe, according to their direct relation 
with the underlying linear line Complex, automatically an 
extended (i.e.~'nonlocal') object -- a set of lines -- in real 
3-space $r$.

\subsection{Some Aspects of Study's work}
\label{sec:studyswork}
Driven by some discussions throughout and following this
year's conferences with respect to Cartan's spin and to
unitary symmetries in QFT, especially when based on one
or more su(2) algebras, we've searched through literature.
So before performing some algebra related to sect.~\ref{sec:geom-r-space},
it is necessary and important to recall few historical 
but obviously forgotten aspects which allows to rearrange
some aspects back into their original context.

Especially when facing the discussion of spin and unitary
symmetries, typically attributed to Pauli, Cartan and su(2)
(and certain su($n$)) algebras, the origins can be easily 
traced back at least to Hesse's transfer principle (1866) 
and Meyer's generalization (1883) (for references and an
outline see \cite{klein:1926}, \S 51) of rational curves 
of a point $\lambda=\tfrac{\lambda_{1}}{\lambda_{2}}$ on 
a line. For $n=2$, we thus obtain (planar) conic sections 
parametrized by three homogeneous coordinates 
$\rho x_{0}=\lambda_{1}^{2}$, $\rho x_{1}=\lambda_{1}\lambda_{2}$
and $\rho x_{2}=\lambda_{2}^{2}$, or appropriate linear 
combinations. Choosing $n=3$ for later use, we obtain 
3$^{rd}$ order curves parametrized by the four coordinates
$\rho x_{0}=\lambda_{1}^{3}$, $\rho x_{1}=\lambda_{1}^{2}\lambda_{2}$, 
$\rho x_{2}=\lambda_{1}\lambda_{2}^{2}$ and $\rho x_{3}=\lambda_{2}^{3}$.
Now linear transformations of the line, alternatively re-expressed
by matrix transformations of the vector/'spinor' $(\lambda_{1}, \lambda_{2})^{T}$
with non-vanishing determinant, i.e.~by a 3-dim group\footnote{German: 
dreigliedrige Gruppe}, lead to collineations in $\mathbbm{R}^{n}$
which exchange points on the (invariant) curve, for $n=2$ the
conic section. For further details see \cite{klein:1926} or
\cite{clebsch:1872}.

In the context of line and Complex geometry -- in a certain 
sense continuing the heritage of \PLL and Lie -- one {\it must}
consider and honor Study's work (see \cite{study:1903}, or 
\cite{study:1907} and references) which has been definitely
known to Cartan. Ad hoc, without enough time to work through
and synchronize notation, we found references and work in 
M{\"u}nchner and Leipziger Berichte (where Cartan published 
in german, too), and Mathematische Annalen, besides Study's 
remarks on Lie's transfer principle in \cite{study:1917}. 
The main task to make the mappings unique will focus on 
introducing 'orientation' so that e.g.~(projective) lines 
correspond to two (oriented) 'Speere' which -- on the 
other hand -- introduces and requires metric (and sometimes
euclidean) concepts and objects, see e.g.~\cite{study:1903}.

Here, mentioning \cite{study:1907}, part II, p.~151, eqns.~(51,l)
and (51,r), Study explains the above 'spinor' notation, and
as such Cartan's 'definition', in 'elliptic' context and 
SU(2)$\times$SU(2) (or quaternionic) transformations, whereas 
in eq.~(52), he resolves the 'spinor' components linearly in 
terms of his 'elliptic' line coordinates.

From our point of view, this emphasizes once more the use of 
Complex geometry. However, Study's part II and his reasoning 
with respect to 'elliptic' geometry has to be regarded with 
time and care in detail, because in eq.~(1) on p.~119 he 
assumes the absolute polar system as 
$(xy)=x_{0}y_{0}+x_{1}y_{1}+x_{2}y_{2}+x_{3}y_{3}$. This allows
him to introduce real combinations of (homogeneous) line coordinates
in his eq.~(3). However, his eq.~(4), re-expressing the \PLL 
condition of the line coordinates $X_{\alpha\beta}$ and using
six line coordinate combinations $X_{i}$, can be cast into 
the Complex expression 
\begin{equation}
X_{1}^{2}-X_{2}^{2}+X_{3}^{2}-X_{4}^{2}+X_{5}^{2}-X_{6}^{2}=0
\end{equation}
while exhibiting SO(3,3) symmetry for a point interpretation
of $X_{i}$, which directly relates to the fundamental form 
of \PLu's line geometry (see also \cite{klein:1926}, \S 23,
especially p.~98).

However, if we switch Study's absolute polar system to 
$(xy)=-x_{0}y_{0}+x_{1}y_{1}+x_{2}y_{2}+x_{3}y_{3}$ by introducing 
an $i$ in the first (point) coordinate $\{\cdot\}_{0}$ so that the
\PLL coordinates $X_{0k}$ change to $iX_{0k}$, the coordinates
(up to overall $i$'s) in Study's eq.~(3) are Klein's coordinates
$z$ (\cite{klein:1926}, \S 23).

Anyhow, citing Klein's work on six Complexe in involution and their
threefold $\pm$-handedness (\cite{klein:1926}, \S 23, p.~99), or 
facing Study's discussion of two spheres (or left/right quaternions)
above, it is evident that these properties represent PG in terms of 
line and Complex geometry which reflects in certain known symmetry 
groups\footnote{U(1) is almost always evident by planar projections.}
like SU(2), SU(2)$\times$SU(2) or even SU(4) of the (spinorial) 
formulation of quantum theories.

\subsection{Basic Spin Commutators}
\label{sect:basiccommutators}
Now, having established the above matrix form of Lie transfer, 
we want to insert a brief section related to practical application
and identification.

This time, we start from the matrix reps above in that we define
points in $R$-space right from the beginning and ask for the 
r\^{o}le of the commutator as natural (antisymmetric) 'product'
of two such objects\footnote{There is some parallel background
from within classical polar theory of second order surfaces, 
however, the calculations here are self-contained so we'll 
discuss those issues in projective geometry at a later time.}.
As such, we define two 'points' ${\cal X}\sim\{X,Y,Z\}$ and
${\cal X}'\sim\{X',Y',Z'\}$, keeping the notation of the 
inhomogeneous coordinates from above, but switching to positive
signs in eqns.~(\ref{eq:spin}) and (\ref{eq:maptopauli}), 
respectively, by
\[
{\cal X}:=X\sigma_{1}+Y\sigma_{2}+Z\sigma_{3}=
\left(
\begin{array}{cc} Z & X-iY\\ X+iY & -Z\end{array}
\right)
\,,\quad
{\cal X}':=
\left(
\begin{array}{cc} Z' & X'-iY'\\ X'+iY' & -Z'\end{array}
\right)\,.
\]
Defining ${\cal C}$ mainly as commutator by
${\cal C}({\cal X},{\cal X}')$:=$\tfrac{1}{2}\left[{\cal X},{\cal X}'\right]$, 
straightforward calculation yields
\begin{equation}
\label{eq:pauliproduct}
{\cal C}=
\left(
\begin{array}{cc}
i(XY'-YX') & ZX'-XZ'+i(YZ'-ZY')\\
XZ'-ZX'+i(YZ'-ZY') & -i(XY'-YX')
\end{array}
\right)\,.
\end{equation}
In order to get more insight, by using the two points 
${\cal X}$ and ${\cal X}'$, we construct line coordinates
in $R$-space, i.e.~we define and use $P_{12}=XY'-YX'$, 
$P_{23}=YZ'-ZY'$ and $P_{31}=ZX'-XZ'$. So eq.~(\ref{eq:pauliproduct})
reads as
\begin{equation}
\label{eq:lineproductR}
{\cal C}=
\left(
\begin{array}{cc}
iP_{12} & P_{31}+iP_{23}\\
-P_{31}+iP_{23} & -iP_{12}
\end{array}
\right)\,=\,
i\left(
\begin{array}{cc}
P_{12} & P_{23}-iP_{31}\\
P_{23}+iP_{31} & -P_{12}
\end{array}
\right)\,,
\end{equation}
where the rhs can be re-expressed in terms of Pauli spin
matrices as ${\cal C}=i(P_{23}\sigma_{1}+P_{31}\sigma_{2}+P_{12}\sigma_{3})$.
Although we are used to having the $i$ accompanying the
commutator of (Pauli) spin matrices, it is easy to transform
eq.~(\ref{eq:lineproductR}) into a real equation by introducing
quaternions, $q_{k}=-i\sigma_{k}$. Eq.~(\ref{eq:lineproductR})
changes into the equation
\begin{equation}
\label{eq:hlplineproduct}
{\cal C}=\frac{1}{2}({\cal X}{\cal X}'-{\cal X}'{\cal X})
=i(P_{23}\sigma_{1}+P_{31}\sigma_{2}+P_{12}\sigma_{3})\,.
\end{equation}
By multiplying both sides with $-1$, i.e.~$(-i)^2$ on the lhs, 
we obtain $-{\cal C}=:\underline{{\cal C}}
=\frac{1}{2}(\underline{X}\,\underline{X}'-\underline{X}'\underline{X})
=\underline{P}$ where each underlining now denotes a real 
quaternion, respectively, e.g.~$\underline{X}=Xq_{1}+Yq_{2}+Zq_{3}$,
$\underline{P}=P_{23}q_{1}+P_{31}q_{2}+P_{12}q_{3})$, etc. 
If we transfer the spin notation above of 
$\underline{{\cal C}}=-{\cal C}(\cal{X},\cal{X}')$
to real quaternion calculus, eq.~(\ref{eq:hlplineproduct}) reads as
$\underline{{\cal C}}(\underline{X},\underline{X}')=\underline{P}$,
i.e.~\emph{the product (or commutator) $\underline{{\cal C}}$ of
two real 'point' quaternions $\underline{X}$ and $\underline{X}'$
yields a real quaternion holding the three spatial \PLL coordinates}.
We may, in addition, recognize the striking index ordering in
eqns.~(\ref{eq:lineproductR}) and (\ref{eq:hlplineproduct}) as
well as ask for further commutators of those objects.

\subsection{Advanced Spin Commutators}
\label{sect:advancedcommutators}
If we denote the components $P_{i}$ of the '\PLu' quaternion 
$\underline{P}$ above through $\underline{P}=P_{i}q_{i}$, by 
performing some algebra we obtain 
$\underline{{\cal C}}(\underline{X},\underline{P})=(P_{kj}x_{j})q_{k}
=:\underline{U}$. Whereas in QFT, we would now have to consider
higher and increasing order of 'vector' products, reducable by
quadratic (mass/momentum) constraints and angular momentum algebra,
here we can stress the origin of $\underline{P}$.

By switching to the general context in terms of homogeneous 
coordinates, we can use the relation of linear Complexe and 
null systems. As such, in terms of homogeneous coordinates
it is $P_{\alpha\beta}$ which relates $x_{\beta}$ to the dual
plane coordinatized by $u_{\alpha}$. Whereas in the polar 
case, incidence $u_{\alpha}x_{\alpha}=0$ of point and plane
is special, treating null systems, the incidence relation is
always satisfied, i.e.~point and plane intersect because of 
antisymmetry of $P_{\alpha\beta}$ in 
$u_{\alpha}x_{\alpha}=P_{\alpha\beta}x_{\beta}x_{\alpha}=0$.

Recalling this background from PG, by considering the real 
quaternion $\underline{U}$ above, its three components realize
(almost) the spatial part of a null system in terms of the 
associated planar coordinates $u_{i}$.

It is obvious that we have to enhance the mechanism because
at this stage we've lost contact to the $0$-components of the
complete null system, and to the $0$-coordinates of points 
and planes in homogeneous rep. What remains open here, is 
the question whether this quaternionic calculus is sufficient
to represent inhomogeneous calculations in mechanics and 
electrodynamics, i.e.~comparison to experimental (classical)
data.

Possible enhancements are obvious because the su(2) rep above
is related to one line only. By recalling the two skew (and 
thus independent) lines above which we can identify either 
as opposite edges of a (fundamental) tetrahedron or as the 
two lines of a Congruency \cite{dahm:MRST4}, we can almost 
automatically enhance the scheme by another independent su(2)
(or quaternion) of the second (skew) line, so that the general
construction scheme comprises SU(2)$\times$SU(2) (or 
SL(1,$\mathbbm{H}$)$\times$SL(1,$\mathbbm{H}$)) or SU(4) (or 
SL(2,$\mathbbm{H}$)$\cong$SU$*$(4)), so we've closed the loop 
to our original starting point SU(4) or Dirac algebra in terms
of SU$*$(4) (see \cite{dahm:MRST4}, \cite{dahm:MRST5} and refs
there). SO(6) and the various real forms SO($n$,$m$), $m+n=6$,
can be related either to their unitary covering groups, or by 
starting from SO(3,3) or SO(6) in line geometry by coordinate 
complexification, either of lines or points (thus, however, 
changing the respective absolute element or polar system, and
in consequence the related underlying geometry). We postpone 
further discussions here.

Last not least, while realizing PG aspects by quaternions and
their commutators, it is important to note that (in general) 
$\underline{{\cal C}}(\underline{X}'',\underline{P})
\sim[\underline{X}'',\underline{P}]
\sim[\underline{X}'',[\underline{X},\underline{X}']]$. This 
discussion can be extended to Jacobi identities, triple systems,
curvature, and further geometric relations.

\subsection{Lagrangean Construction}
\label{sec:lagrangean}
So far, after having related 'spinors' and SU(2) spin reps
by means of Lie transfer to line and Complex geometry, we 
have to focus on appropriate Lagrangean constructions in 
order to reflect Complex invariants and dynamics, and  
provide the Lagrangean (or Hamiltonian) apparatus.

Here, we'll firstly focus on the photon rep of the QED 
Lagrangean, i.e.~we'll postpone the construction of matter
fields and the discussion of Dirac or Clifford algebra for
a moment and focus on the construction principle using Complexe.
This reflects the fact\footnote{For details, see \cite{dahm:MRST5}
sec.~III with respect to relativistic symmetry and to symplectic
geometry, thus stabilizing a certain linear Complex.} that 
the photon rep almost automatically corresponds to a special
linear Complex \cite{dahm:MRST4}, \cite{dahm:MRST5} while 
obeying (and respecting) special relativity. Relativistic 
aspects will be treated in more detail in part VIII.

Above, we've shown that the 'spinor' discussion is a subset 
of Complex geometry right from Lie's paper. So to discuss 
invariance and construct invariant elements in Complex geometry,
we have lots of geometrical possibilities, some of which 
we've already mentioned elsewhere. On the ground of Klein's
Erlanger program, being in the regime of projective transformations,
we may stress directly (quadratic) tetrahedral Complexe leaving 
anharmonic ratios of line intersections with a (fundamental) 
tetrahedron invariant. With respect to second order surfaces 
to be discussed in part VIII as absolute elements of non-euclidean
geometry, we may as well consider polar or tangential Complexe. 
And possible degeneracy or Clebsch's paper on Complex symbolism 
\cite{clebsch:1869} suggests to consider assemblies or powers 
of individual (special) linear Complexe as well in the basic
construction process.

Now the first major step is to identify electromagnetism by means
of a special linear Complex. The \PLL condition 
$P:=p_{01}p_{23}+p_{02}p_{31}+p_{03}p_{12}=0$ suggests to identify 
the field strengths $\vec{E}$ and $\vec{B}$, i.e.~we have 
$\vec{E}\cdot\vec{B}=0$, so initially we can identify $\vec{E}_{i}$
and $\vec{B}_{i}$ as six \PLL coordinates. Having argued in 
\cite{dahm:MRST5} with Klein's papers of 1869 and 1871 (see also
\cite{klein:1926}), here we introduce some linear combinations 
$y$ of \PLu's line coordinates $p\sim x$ according to the notation
\cite{klein:1926}, p.~98, and we obtain (up to a factor 4)
\begin{equation}
\vec{E}^{2}-\vec{B}^{2}\,=\,\vec{E}^{2}+\vec{iB}^{2}
\,=:\,\Omega\,,
\end{equation}
which Klein defines as an invariant $\Omega$ of the linear Complex
and states $\Omega=0$ as condition for a special linear Complex.
Recall that this decomposition historically gave rise to various
discussions of SU(2)$\times$SU(2) reps, as well as various 
complexifications thereof, mostly celebrating vector/'spinor'
calculus.

Re-expressed in contemporary notation, we can compare e.g.~with
\[
-F_{\mu\nu}F^{\mu\nu}\sim\vec{E}^{2}-\vec{B}^{2}\,=\,\Omega\,,\quad
G_{\mu\nu}F^{\mu\nu}\sim\vec{E}\cdot\vec{B}=\,P\,,\,
G_{\mu\nu}=\frac{1}{2}\epsilon_{\alpha\beta\mu\nu}F^{\alpha\beta}
\]
being the hodge dual which we've related earlier to conjugation 
and polar theory of Complexe.

So working with simple invariants of linear special Complexe,
only, comprises known QED Lagrangean formulations. From the 
decomposition of quadratic Complexe or investigating pencils
of linear Complexe, we know that we have to treat regular 
Complexe as well, so besides using the above quadratic expressions
to enforce projective invariance or absolute elements, we expect 
the individual as well as the simultaneous invariance theory 
of general linear Complexe (special as well as regular) to 
comprise an appropriate basis of Lagrangean construction. 
Standard unitary symmetries are included as we've shown already
above.

\section{Summary and Outlook}
\label{sec:outlook}
Essentially, we have shown that su(2) and the Pauli rep according
to Lie's setup is an artifact emerging from line geometry, and a 
degenerated case of Complex geometry. Lie transfer puts classical
PG and sphere geometry side by side, and we've found especially 
real quaternion calculus formally realizing known geometry based
on null systems, as soon as one understands the commutator as a 
generalized product. However, this needs deeper and more complete
research. The general framework superseding this type of algebra 
is Complex geometry, and we have connected previous work by closing
the circle to twofold quaternions and certain unitary symmetries 
we've started with. The outline and development, however, suggests
to consider typical spin applications or SU(2) symmetry in physics 
from the viewpoint of line or Complex geometry as well.

\end{document}